\documentclass[aps,superscriptaddress,floats,showpacs,twocolumn]{revtex4}
\usepackage{graphicx}
\usepackage{subfigure}
\usepackage{bm}
\begin{document}
\title{Geometry of intensive scalar dissipation events in turbulence}
\author{Dan Kushnir}
\affiliation{Department of Computer Science and Applied Mathematics,
           The Weizmann Institute of Science, 76100 Rehovot, Israel}
\author{J\"org Schumacher}
\affiliation{Department of Mechanical Engineering,
           Technische Universit\"at Ilmenau, D-98684 Ilmenau, Germany}
\author{Achi Brandt}
\affiliation{Department of Computer Science and Applied Mathematics,
           The Weizmann Institute of Science, 76100 Rehovot, Israel}
\date{\today}
\begin{abstract}
Maxima of the scalar dissipation rate in turbulence appear in form of sheets and
correspond to the potentially most intensive scalar mixing events. 
Their cross-section extension determines a locally varying diffusion scale of the mixing process
and extends the classical Batchelor picture of one mean diffusion scale. 
The distribution of the local diffusion scales is analysed for different Reynolds and Schmidt numbers with a 
fast multiscale technique applied to very high-resolution simulation data. The scales take always
values across the whole Batchelor range and beyond. Furthermore, their distribution 
is traced back to the distribution of the contractive short-time Lyapunov exponent of the flow. 
\end{abstract}
\maketitle

When a scalar concentration field $\theta({\bf x},t)$ is transported in a 
turbulent flow very large scalar gradients are generated which can be 
associated with
potentially intensive mixing \cite{Dimotakis2005}. 
Frequently, such large-amplitude gradient 
regions exist across scales that are finer than the smallest turbulent eddies, a case
which is known as the 
Batchelor regime of scalar mixing \cite{Batchelor1959}. Their cross-section is  
usually estimated by a single mean diffusion scale, the Batchelor scale $\eta_B$, that equilibrates 
advection by flow
and scalar diffusion. However, scalar gradients are known to fluctuate 
strongly in turbulent mixing. These fluctuations are caused by the fluctuating scalar amplitudes 
and by the varying spatial sections across which the scalar differences are built up. 
Both aspects will cause the strong 
spatial variability of potentially intensive mixing regions. Thus, a whole range of {\em local} 
diffusion scales $l_d$, which quantifies exactly this variability, will exist around the mean diffusion scale
$\eta_B$. Their distribution is interesting for several reasons. It can enter mixing 
efficiency measures \cite{Doering2004}. The scales prescribe the extension of  
chemically reactive layers in which the combustion of fuel takes place \cite{Peters2000} or the 
variations of the fluorescence signal as used for the measurement of zooplankton patchiness in the 
ocean \cite{Wolk}. 

In this Letter, we want to calculate this local diffusion scale distribution $p(l_d)$. 
It arises from the competition of two dynamic processes. On one hand, the scale distribution 
will be affected by molecular diffusion that causes a diminishing of existing steep
gradients as well as the completion of their formation by
reconnection \cite{Gibson1968,Villermaux2003}. On the other hand, the
scales  will be determined by the statistics of the local advection flow patterns that pile 
up scalar differences. While  
the strongest scalar gradients were related to instantaneous velocity gradients in 
Ref.~\cite{Schumacher2003}, we will go one step further in the second part of the Letter and relate 
the local diffusion scale 
distribution to the distribution of Lagrangian contraction rates of the flow, as given by 
the smallest of the three finite-time Lyapunov exponents along Lagrangian trajectories. 
The Lagrangian approach incorporates 
the temporal evolution of the persistent flow patterns that eventually generate the strongest scalar 
gradients or the finest local dissipation scales in a finite time.       
The analysis is conducted for passive scalars in Navier-Stokes turbulence, a case 
that is not accessible to analytical treatment. Our investigations are therefore based on direct 
numerical simulations (DNS). In order to discuss trends with Reynolds and Schmidt number, 
we analyze five different data sets. 
Very high-resolution calculations (with up to $1024^3$ grid points) have to be conducted with 
resolution constraints that exceed the usually applied ones such that 
the finest scales of the turbulent mixing process are resolved properly. The fluctuations of the
flow are sustained statistically stationary by a large-scale random forcing and the passive scalar is driven by
a constant mean scalar gradient. We consider the scalar 
dissipation field \cite{diss} which probes the magnitude of scalar gradients and which is defined as
\begin{equation}
\epsilon_{\theta}({\bf x},t)=\kappa |{\bf\nabla}\theta({\bf x},t)|^2\,.
\end{equation}
Here, $\kappa$ is the scalar diffusivity. 
Experimental studies on the geometry of scalar dissipation fields 
are very challenging since gradients have to be measured and only a few exist \cite{Buch1998,Su2003}. 
  
Figure~1 shows a two-dimensional (2D) slice cut through a DNS snapshot of 
$\epsilon_{\theta}({\bf x},t)$. We observe strongly folded filaments. The points that form the 
level set of largest dissipation amplitudes
\begin{equation}
L_C=\{ {\bf x}:\epsilon_{\theta}\ge C
\langle\epsilon_{\theta}\rangle\; \mbox{or}\; |{\bf\nabla}\theta|\ge 
\sqrt{C \langle|\nabla\theta|^2\rangle}\}\,,
\label{level}
\end{equation}
\begin{figure}
\centerline{\includegraphics[angle=0,scale=0.18,draft=false]{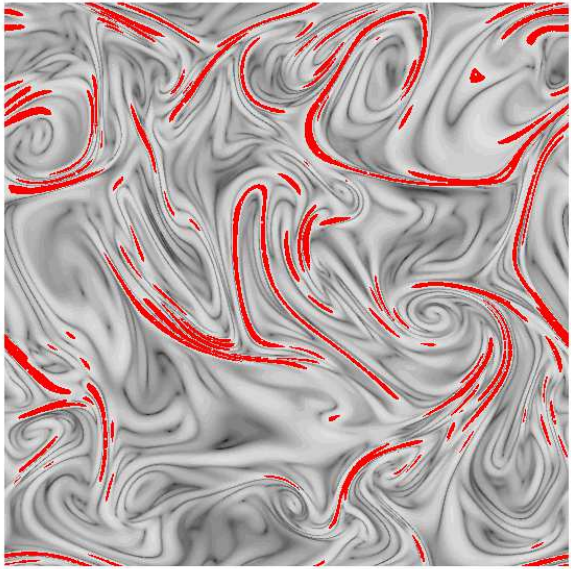}}
\caption{(color online) Contour plot of a two-dimensional slice
cut through the instantaneous three-dimensional scalar dissipation
rate field. Level set $L_C$ for
$C=4$ (see Eq.~(\ref{level})) is replotted in red. Data are from a
pseudospectral simulation of the advection-diffusion equation for
the passive scalar in combination with Navier-Stokes equations for
a statistically stationary, homogeneous isotropic flow at a
resolution of $1024^3$ grid points in a periodic box 
$V=(2\pi)^3$ \cite{Schumacher2003}.  The
Schmidt number is $Sc=\nu/\kappa=32$ and the Taylor microscale Reynolds
number is $R_{\lambda}=\sqrt{15/(\nu\langle
\epsilon\rangle)}\,\langle u_x^2\rangle=24$ with
$\langle\epsilon\rangle$ being the mean energy dissipation rate and $\nu$ the
kinematic viscosity.
The Batchelor scale
$\eta_B=\eta/\sqrt{Sc}$ is resolved with 2 grid cells.
The viscous Kolmogorov length $\eta=\nu^{3/4}/\langle\epsilon\rangle^{1/4}$ measures 
then 11 grid cells.
This spectral resolution is larger by a factor of 4
than the one usually adopted.} \label{sheets}
\end{figure}
are redrawn in red. $C$ is a real constant.
The resulting filaments are cross-sections of
thin sheets in which the maxima of scalar dissipation are arranged in
the three-dimensional volume \cite{Schumacher2003}. A closer
inspection of Fig.~\ref{sheets} unravels various length and
thickness scales of the filaments. The filaments are curved and
tightly clustered in certain locations thus posing a challenge
of separating each curved filament and computing its accurate
variation scales.  

\begin{figure}
\centerline{\includegraphics[angle=0,scale=0.3,draft=false]{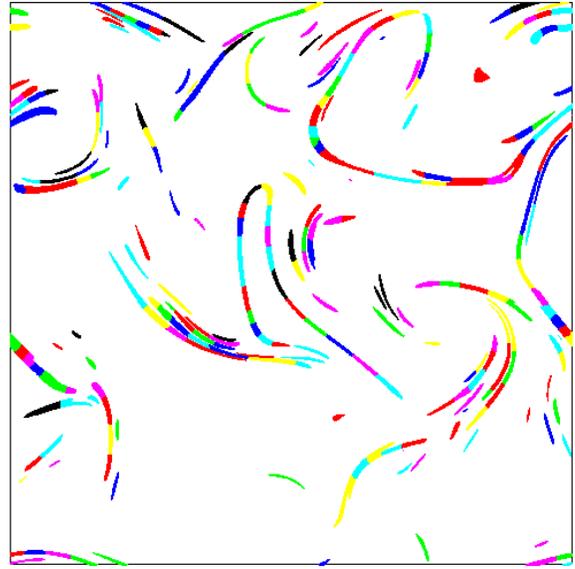}}
\caption{(color online)
Reconstruction of the red colored filaments as shown in Fig. 1 by means of the
fast multiscale clustering algorithm. Long filaments are composed of several
subfilaments that are colored differently.}
\label{aggregates}
\end{figure}
This is done here by a
fast multiscale clustering algorithm \cite{Brandt2004}, which is based on the 
Segmentation by Weighted
Aggregation (SWA) algorithm \cite{AMG}, motivated by Algebraic
Multigrid (AMG) \cite{SharonBrandtBasri2000}. The algorithm
assigns data points into clusters starting at the finest resolution level, the
grid spacing. All points ${\bf x}_i\in L_C$ are therefore gathered in a
so-called proximity graph that contains their location and their
connectivity to other points of $L_C$ as quantified by the
inter-point weights which probe the nearest graph neighbors of each point only. Since the
information about proximity is kept as one moves from finer to
coarser resolution level, one can perform a fast recursive
principal component analysis (PCA). The resulting eigenvalues 
characterize the length and width of the point clusters. Additionally, strongly curved filaments
have to be decomposed into sub-filaments by applying a local convexity 
criterion along the filament. Figure~\ref{aggregates} shows the
reconstruction of the filaments from Fig.~1 and their division
into sub-filaments. The local dissipation filament thickness,
$l_d$, is then given by nothing else but the smaller eigenvalue
which follows from the PCA of each sub-filament.
\begin{figure}
\centerline{\includegraphics[angle=0,scale=0.6,draft=false]{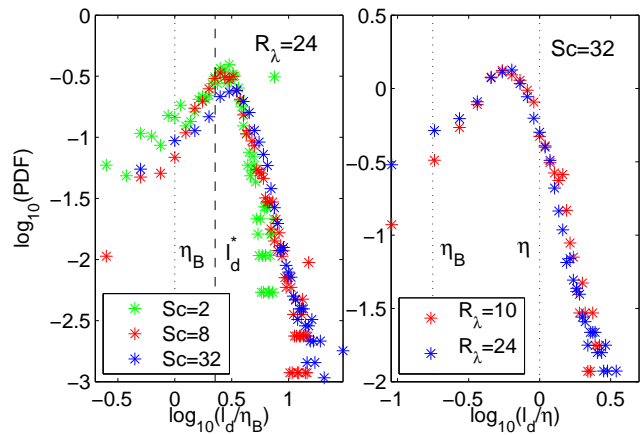}}
\caption{(color online)
Distribution of the local cross-section thickness $l_d$ of the scalar dissipation
rate filaments for $\epsilon_{\theta}\ge 4\langle\epsilon_{\theta}\rangle$.
Left panel: Probability density function (PDF) $p(l_d/\eta_B)$ for three different
Schmidt numbers at $R_{\lambda}=24$. The dashed line corresponds with the theoretical 
value of the most probable scale $l_d^*=\sqrt{\kappa/|\Lambda_3|}$ which will be discussed later in the text.
Right panel: PDF
$p(l_d/\eta)$ for two different Reynolds numbers at $Sc=32$.}
\label{hist}
\end{figure}

Figure~\ref{hist} shows the probability density functions (PDF) of
the local filament thickness $l_d$ for different Schmidt ($Sc$) and Reynolds numbers ($R_{\lambda}$). 
The distribution will depend on the cut-off level
$C$, but the physical picture will not change since $C$ is fixed with
respect to the mean scalar dissipation rate throughout
the analysis. Local thickness values within
the whole Batchelor range between $\eta_B$ and $\eta$ and beyond are found,
indicating that dissipation maxima are also related to scalar gradients
across inertial range scales. The 2D analysis does not account for the
spatial orientation of the sheets with respect to the cutting
plane. As demonstrated in \cite{Buch1998}, this will affect only
the tail for large $l_d$. The left panel of Fig.~\ref{hist}
compares the PDFs for three different Schmidt numbers at a fixed
Reynolds number $R_{\lambda}=24$. With increasing Schmidt number stronger jumps of
the scalar concentration across finer thickness scales become more
probable since diffusion is less dominant. Consequently, the most
probable thickness $l_{max}$, i.e. the maximum of the distribution, 
is shifted to smaller values,
but remains always larger than the corresponding
Batchelor scale. This suggests that the formation of so-called
mature scalar gradient fronts with a thickness $\sim\eta_B$ is a
subdominant process. We see that the three PDFs overlap when rescaled
with $\eta_B$ which implies that $l_{max}\sim Sc^{-1/2}$ on the basis of our data and
even though $Sc=2$ has no real Batchelor range. In
the right panel of Fig.~\ref{hist}, we compare distributions for
two different Reynolds numbers at fixed Schmidt number. Both PDFs
rescaled with the corresponding $\eta$ overlap again to a large
fraction, except for the very fine scales. Their higher
probability with increasing $R_{\lambda}$ indicates a more
efficient stirring at the smallest scales. Our data show that the scale $l_{max}$
follows now the same dependence with Reynolds number as the
Kolmogorov scale, i.e. $l_{max}\sim R_{\lambda}^{-3/2}$
This result is similar to finding in \cite{Buch1998,Su2003} for $Sc\sim 1$ and does not change 
for $Sc\gg 1$.

\begin{figure}
\centerline{\includegraphics[angle=0,scale=0.6,draft=false]{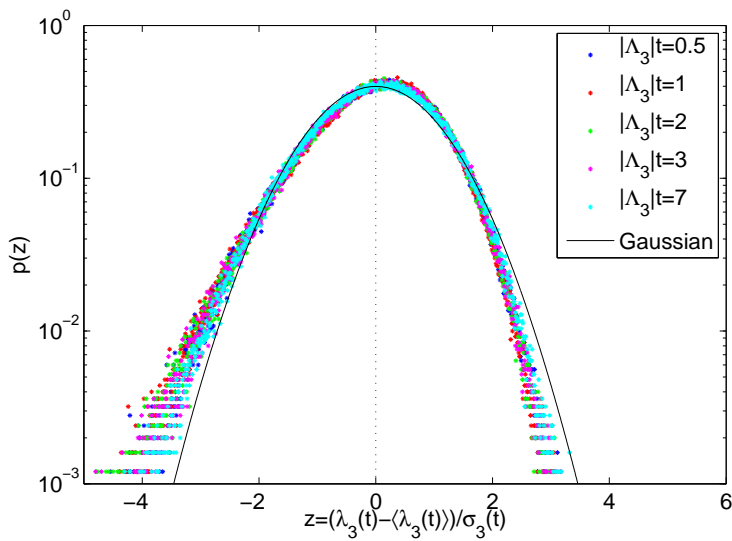}}
\caption{(color online)
Probability density function of $\lambda_3(t)$
for different times. The data are for $R_{\lambda}=24$. The variable
$z=(\lambda_3(t)-\langle\lambda_3(t)\rangle)/\sigma_3(t)$ with
$\sigma_3(t)=\sqrt{\langle (\lambda_3(t)-\langle\lambda_3(t)\rangle)^2\rangle}$ is chosen
for comparison with a Gaussian distribution.
We found $\Lambda_3\simeq -0.425$ and $\sigma^2_3(t)=\sigma^2_{\infty}/t$ with
$\sigma^2_{\infty}\simeq 0.44$ for $t\gtrsim|\Lambda_3|^{-1}$. The distribution is evaluated by
$2.5\times 10^5$ Lagrangian tracers that are initially seeded uniformly. The distribution for
$R_{\lambda}=10$ behaves qualitatively the same.}
\label{double0}
\end{figure}

The filament thickness distribution is now related to
the distribution of finite-time Lyapunov exponents (FTLE) $\lambda_i(t)$. 
They measure the local separation between
two initially infinitesimally close fluid elements along the Lagrangian trajectory.
Therefore an orthogonal frame is attached to each tracer and the three separation vectors
evolve as $\mbox{d}\delta r_j^{(i)}(t)/\mbox{d}t=\sigma_{jk}(t)\,\delta r_k^{(i)}(t)$ for 
$j,k=x,y,z$ and $i=1,2,3$. $\sigma_{jk}(t)$ is the rate of strain tensor along 
the Lagrangian trajectories.
The FTLEs follow to $\lambda_i(t)^=1/t\,\log(|\delta {\bf r}^{(i)}(t)|/|\delta {\bf r}^{(i)}(0)|)$ 
where the Gram-Schmid method is applied consecutively to three $\delta {\bf r}^{(i)}(t)$ 
(see Appendix C.3 of \cite{Bohr1998}). The exponents will
vary from trajectory to trajectory resulting in statistical distributions. 
Their means are the global FTLE, $\langle\lambda_i(t)\rangle$, which
converge for long times to three numbers, the asymptotic Lyapunov exponents 
$\Lambda_i=\mbox{lim}_{t\to\infty}\langle\lambda_i(t)\rangle$.
Due to incompressibility, $\sum_{i=1}^3\lambda_i(t)=0$. Our interest is in the
formation of thin dissipation (or gradient) sheets where expansion in two directions is present,
i.e. $\lambda_1(t)>0$ and $\lambda_2(t)>0$, and contraction in the third one, $\lambda_3(t)<0$.
The distribution of contraction rates that pile up scalar gradient maxima and form the 
distribution of the local diffusion scales follows consequently from the
PDF $p(\lambda_3(t))$ as shown in Fig.~\ref{double0}. We see that the cores of the distributions 
for different times collapse to a Gaussian profile within $\pm 2\sigma_3(t)$.
Since the standard deviation
$\sigma_3(t)=\sigma_{\infty}/\sqrt{t}$ for larger times the contraction
rate $\lambda_3(t)$ will get more and more concentrated about $\Lambda_3$ as the time advances.
Based on $\Lambda_3$ the most probable thickness is given as
$l_d^{\ast}=\sqrt{\kappa/|\Lambda_3|}$ \cite{Balkovsky1999} which arises by equilibrating contractive
strain and diffusion. For all data analysed this scale is at about the maximum of the thickness
distribution, $l_d^{\ast}\simeq l_{max}$ (see left panel of Fig.~3). 

Hu and Pierrehumbert \cite{Pierrehumbert2002} pointed out that the asymptotic $\Lambda_3$ alone is not
sufficient to explain the formation of the finest diffusion
scales when the flow is time-correlated as being the case for the present studies. 
The time that a scalar blob or filament experiences a persistent strain pattern along the trajectory 
can be estimated by a characteristic decorrelation time which is given by $\tau_c\sim \Lambda_3^{-1}$. 
For larger periods the persistent compression of the blob will disappear or the sheets undergo
diffusive destruction or merging. This local scenario will appear repeatedly and causes a stationary
thickness scale distribution. 
Consequently, we have to take into account the evolution of distribution of the FTLEs over such 
periods in order to study the formation of the gradient (or dissipation) sheets. 
The distribution of the sheets is related to the characteristic distribution of the FTLE over 
times $t\lesssim\tau_c$ by 
\begin{eqnarray}
p(l_d)&\sim&\int \mbox{d}l_0\, \tilde{p}(l_0)
          \int_{-\infty}^{+\infty} \mbox{d}\lambda_3\, g(\lambda_3)\,
          \delta\left(l_d-l_0\mbox{e}^{\lambda_3 \tau_c}\right)\nonumber\\
        &=&\frac{1}{l_d \tau_c}\int \mbox{d}l_0\, \tilde{p}(l_0)\,g(\log(l_d/\l_0)/\tau_c)\,,
\label{recons1}
\end{eqnarray}
\begin{figure}
\centerline{\includegraphics[angle=0,scale=0.6,draft=false]{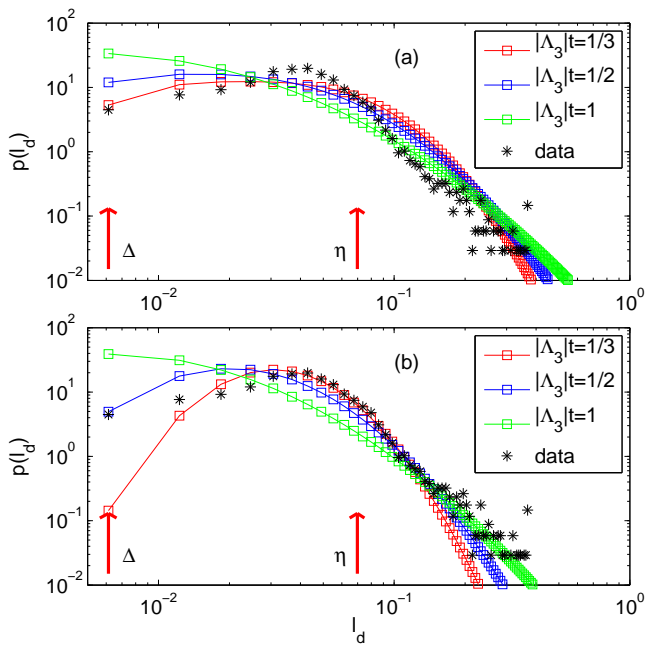}}
\caption{(color online)
Distribution of the local cross-section thickness $l_d$ as reconstructed from
the distribution of the FTLE $g(\lambda_3)$ via Eq.~(\ref{recons1}) for different times. 
Data are for $R_{\lambda}=24$ and $Sc=32$ (see Fig.~\ref{hist}) . 
Two initial distributions are shown:  (a) uniform distribution $\tilde{p}(l_0)=1/(2\eta_K-\eta_B)$;
(b) delta function $\tilde{p}(l_0)=\delta(l_0-1.8 l_d^*)$ (see Fig.~3). The grid spacing $\Delta$
and Kolmogorov scale $\eta$ are indicated as vertical arrows.}
\label{double}
\end{figure}
where the distribution of the FTLE is approximated by a Gaussian $g(\lambda_3,t)=(1/\sqrt{2\pi\sigma_3(t)^2}) 
\exp[-(\lambda_3(t)-\langle\lambda_3(t)\rangle)^2/
(2\sigma_3(t)^2)]$
and $\tilde{p}(l_0)$ the (unknown) distribution of the initial thickness scales.
The initial distributions $\tilde{p}(l_0)$ taken here mark both ends of a spectrum of possible choices,
the localized case given by a delta-function around the maximum of $p(l_d)$ and a uniform distribution
of initial scales over an interval, respectively. 
Both cases can be carried out analytically and result in scale distributions of log-normal type. 
Figure 5 shows the resulting PDFs for $R_{\lambda}=24$ and
$Sc=32$. The distributions agree well with the data for most $l_d$ and for time periods 
$t \lesssim |\Lambda_3|^{-1}$. The uniform case fits slightly better for the left tail since
sufficiently small scales are present initially that can be steepened further to smaller cross 
sections. The strain which is accumulated over short times only seems to explain
the formation of the finest scalar gradient sheets. 
Nearly, the same time scales and distributions were found for the other Schmidt and 
Reynolds number values. 

In conclusion, we have determined the distribution of the cross-section 
extensions of the scalar dissipation maxima in Navier-Stokes turbulence.
These scales correspond with local diffusion scales and take values across the 
whole Batchelor range and beyond. By means of the distribution of smallest 
short-time Lyapunov exponent, the diffusion scale distribution can be reconstructed. 

{\em Acknowledgements.}
We acknowledge support by the NSF grant DMS-9810282 and
thank the Institute for Pure and Applied Mathematics at UCLA for hospitality during
the Multiscale Geometric Analysis program. J.S. was also supported by the DFG.
The simulations were done on the
JUMP supercomputer at the John von Neumann-Institute for Computing of the Forschungszentrum J\"ulich.
We thank W.J.A. Dahm, F. de Lillo, J. Davoudi, B. Eckhardt, P.W. Jones, and A. Thess for helpful 
comments.

\end{document}